\documentclass[letter,twocolumn]{jpsj2}

\title{Impurity Effect on Superconducting Properties in Molecular Substituted \\
Organic Superconductor $\kappa$-(ET)$_2$Cu(NCS)$_2$}

\author{Naoki \textsc{YONEYAMA}\thanks{E-mail address: 
yone@imr.tohoku.ac.jp}, Takahiko \textsc{SASAKI},
Hajime \textsc{OIZUMI} and Norio \textsc{KOBAYASHI}} 

\inst{$^{1}$Institute for Materials Research, Tohoku University, Sendai
980-8577, Japan}

\abst{
We report an impurity effect in the organic superconductor
$\kappa$-(ET)$_2$Cu(NCS)$_2$ by substitution of the 
ET molecule with an analogue, bis(methyleneditio)tetrathiafulvalene (MT).
The superconducting transition temperature decreases with increasing
substitution.
The in-plane magnetic penetration depth is enhanced with 
substitution, which is quantitatively attributed to the decrease in the 
in-plane mean free path.
The enhancement of the penetration depth can also explain the reduction of
the effective pinning in terms of the condensation energy.
}

\kword{organic superconductor, impurity, penetration depth,
 mean free path, molecular substitution, magnetization}

\begin{document}
\maketitle


Organic superconductors $\kappa$-(ET)$_2X$ (ET denotes 
bis(ethylenedithio)tetrathiafulvalene, $X$ = Cu(NCS)$_2$ and 
Cu[N(CN)$_2$]Br)
consist of alternating layers of conducting ET donors and insulating $X$ 
anions.\cite{Ishiguro}
Reflecting the layered structure, the system is well characterized by
highly anisotropic (quasi-two-dimensional) electronic properties.
A prominent feature of the organic superconductors is its clean system 
with less disorder.
In $\kappa$-(ET)$_2$Cu(NCS)$_2$, one of the most vigorously investigated
organics, the in-plane mean free path $l_{\parallel}$ grows to the 
length of approximately 150 -- 200 nm.\cite{141,159}
Taking account of its long in-plane penetration depth 
($\lambda_{\parallel}\sim0.5\ \mu$m)
and short in-plane coherence length ($\xi_{\parallel} \sim\ 5$ nm), the local clean limit
approximation\cite{Tinkham} ($\xi_{\parallel} \ll l_{\parallel} \ll \lambda_{\parallel}$) can be 
properly applied to the superconducting state.

It has been known that strong electron correlations play an
important role on the anomalous superconducting properties, such as 
existence of an antiferromagnetic phase next to the superconductivity.\cite{1106}
This is presumably reminiscent of unconventional superconductivity in 
high-$T_{\textrm{c}}$ cuprates, in which a $d$-wave symmetry of the order 
parameter has been widely accepted.
However in organic superconductors, 
although many efforts have been achieved and its unconventional
($d$-wave) symmetry may be predominantly accepted,
no clear consensus has been reached yet.\cite{977}

To identify the symmetry of the order parameter,
a very helpful method would be to investigate impurity effects.
There have been many studies to introduce impurity (or disorder) into organics:
anion\cite{1421,1425,1149,1426} or donor \cite{1422,1423,1179,1210,1424} 
substitution, fast electron\cite{1427} or X-ray\cite{1379} irradiation, 
and fast-cooling\cite{146}. 
As expected, they commonly suppress $T_{\textrm{c}}$ with increasing disorder
regardless of its origin.
A fully deuterated ET donor (dET) substituted salt
$\kappa$-[(ET)$_{1-x}$(dET)$_x$]$_2$Cu(NCS)$_2$
is a good example of very weakly disordered systems, 
in which $T_{\textrm{c}}$ \textit{increases} with $x$ because of chemical
pressure\cite{1210,1424}.
Nevertheless, the scattering time barely takes minimum at around 
$x=0.5$, indicating that the substitution gives a small
scattering.\cite{1424}
For more detailed investigation, a systematic introduction and control 
of stronger disorder would be indispensable.

\begin{figure}
\begin{center}
\includegraphics[clip,width=7cm]{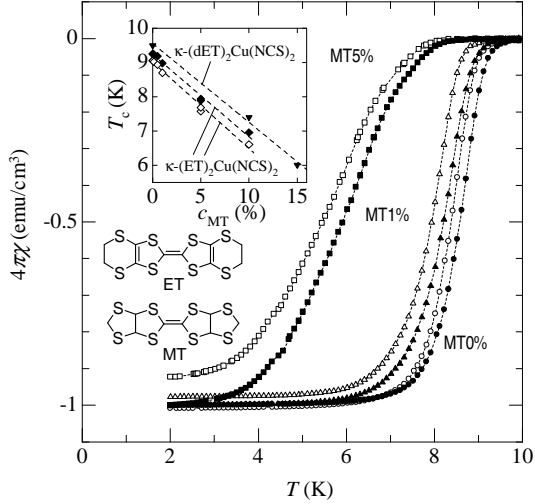}
\caption{\label{Fig:kaiT}Temperature dependence of the susceptibility
of the MT substituted $\kappa$-(ET)$_2$Cu(NCS)$_2$.
The data of MT0\% were taken from the previous work.\cite{1151}
Filled and open symbols are obtained in the slow-cooled and quenched
conditions, respectively.
The upper inset shows $T_{\textrm{c}}$ as a function of the concentration of MT.
The filled (open) diamonds denote $\kappa$-(ET)$_2$Cu(NCS)$_2$ in 
slow-cooled (quenched) condition, and inversed triangles the slow-cooled 
data in $\kappa$-(dET)$_2$Cu(NCS)$_2$.
Broken lines are guides for the eye.
The lower inset displays ET and MT molecules.
}
\end{center}
 \end{figure}

In this paper, we present the substitution of a relative
donor molecule bis(methylenedithio)tetrathiafulvalene (abbreviated as MT 
or BMDT-TTF, see lower inset of Fig. \ref{Fig:kaiT}) for the ET molecule.
In the substituted samples, $T_{\textrm{c}}$ monotonically decreases with increasing substitution.
This substitution enhances in-plane magnetic penetration depth, 
which is interpreted as an impurity effect 
on the basis of the clean local limit approximation.

Single crystals of the MT substituted $\kappa$-(ET)$_2$Cu(NCS)$_2$ were 
grown by a standard electrochemical technique.
We label the samples using the MT concentration $c_{\textrm{MT}}$ as,
for example, ``MT1\%'', which was prepared with 1 mg of MT and 99mg of ET
dissolved into 100 mL 1,1,2-trichloroethane (5\% ethanol).
Samples of MT0\%, MT0.01\%, MT0.1\%, MT0.5\%, MT1\%, MT5\%, MT10\%, and MT15\%
were prepared.
The crystalline shape and quality of the substituted samples were almost
the same ones with the pristine MT0\% sample.
The magnetization measurements were performed using a SQUID magnetometer 
(Quantum Design, MPMS-XL).
Each sample was cooled with rates of approximately $-100$ K/min from room temperature 
(quenched) and $-0.2$ K/min from 100 K (slow-cooled).
Temperature and magnetic field dependences of the magnetization were 
measured in zero field cooled condition.
All the data for MT0\% shown here were taken from the previous study.\cite{1151}

First of all, to quantify the amount of the actual substitution, we carried out
FT-IR spectroscopy in $\kappa$-(dET)$_2$Cu(NCS)$_2$ at
$c_{\textrm{MT}}=5, 10,$ and 15\% (data not shown in the figure).
A peak intensity of the molecular vibration mode in terminal ethylene 
groups has been used to determine the substitution ratio in 
$\kappa$-[(ET)$_{1-x}$(dET)$_x$]$_2$Cu[N(CN)$_2$]Br,\cite{1308} since the peak
height of the vibration mode linearly alters with $x$.
In the present case as well, the peak height in the pristine $\kappa$-(dET)$_2$Cu(NCS)$_2$
reduces to approximately 85$\pm4$\% in MT15\%, 
although no significant variation is observed at more dilute 
$c_{\textrm{MT}}$ within experimental error.
Taking into account the linear decrease in $T_{\textrm{c}}$ with $c_{\textrm{MT}}$ 
as described below, 
it is concluded that the actual value of the MT concentration is 
in the same order with $c_{\textrm{MT}}$ in the whole concentration 
prepared in the present work.

\begin{figure}
\begin{center}
\includegraphics[clip,width=7cm]{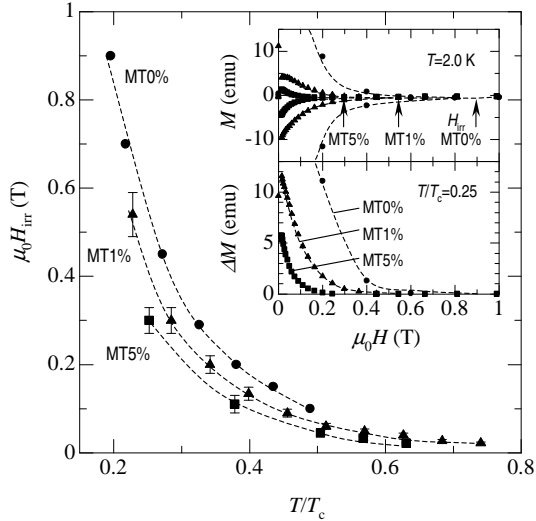}
\caption{\label{Fig:Hirr}Irreversibility field as a function of $T/T_{\textrm{c}}$ in 
the MT substituted $\kappa$-(ET)$_2$Cu(NCS)$_2$ in the slow-cooled 
condition for MT0\% (circle), MT1\% (triangle), and MT5\% (square).
The upper inset shows the magnetization curves at 2.0 K, and the lower 
one the magnetization hysteresis widths at $T/T_{\textrm{c}}=0.25$.
Broken curves are guides for the eye.
}
\end{center}
\end{figure}

\begin{figure}
\begin{center}
\includegraphics[clip,width=7cm]{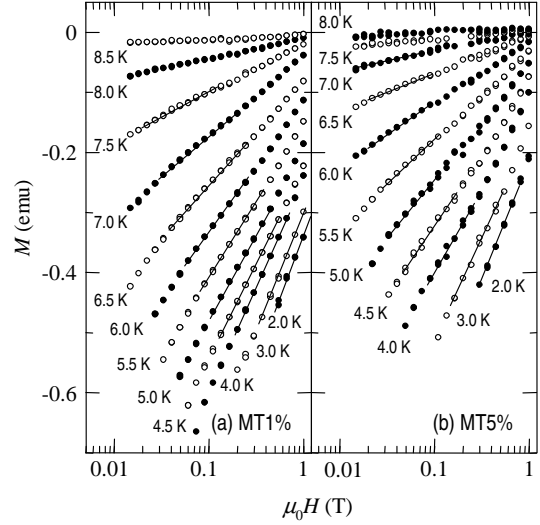}
 \caption{\label{Fig:MlnH} Semi-logarithmic plots of the reversible magnetization 
 curves as a function of magnetic field in the MT substituted 
 $\kappa$-(ET)$_2$Cu(NCS)$_2$ (a) in MT1\% and (b) MT5\%, respectively.
 Data were taken in the slow-cooled condition.
 Data points below $H_{\textrm{irr}}$ were omitted for clarity.
 Solid lines indicate the linear regions used to evaluate $\lambda_{\parallel}$.
}
\end{center}
\end{figure}

\begin{figure}
\begin{center}
\includegraphics[clip,width=7cm]{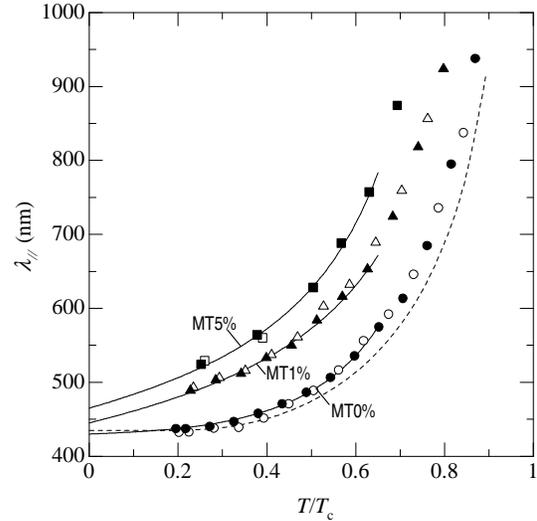}
 \caption{\label{Fig:lambda} In-plane magnetic penetration depth of 
the MT substituted $\kappa$-(ET)$_2$Cu(NCS)$_2$ at $c_{\textrm{MT}}=0$\% (circles), 
1\% (triangles), and 5\% (squares).
The filled (open) symbols indicate the data in slow-cooled (quenched) condition.
Solid curves are obtained on the basis of a $d$-wave model (see text), 
and a broken curve is an s-wave (local clean) model.
}
\end{center}

\end{figure}
\begin{figure}
\begin{center}
\includegraphics[clip,width=7cm]{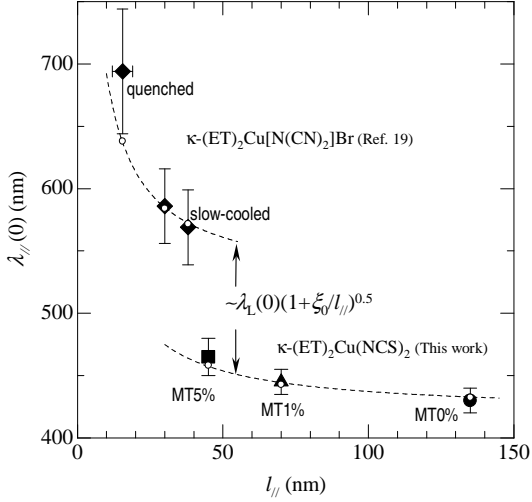}
 \caption{\label{Fig:London} In-plane magnetic penetration depth as a 
 function of the mean free path in $\kappa$-(ET)$_2X$ ($X$=Cu(NCS)$_2$, Cu[N(CN)$_2$]Br). 
The data for the latter one are taken from the literature.\cite{1151}
Small open circles are calculated from eq. (\ref{Eq:London}) with 
fitting parameters of $\lambda_{\textrm{L}}(0)=420$ and 523 nm 
for $X=$Cu(NCS)$_2$ and Cu[N(CN)$_2$]Br, respectively.
Broken curves are guides for the eye.
}
\end{center}
\end{figure}


Figure\ \ref{Fig:kaiT} shows the temperature dependence of the susceptibility in 
$\kappa$-(ET)$_2$Cu(NCS)$_2$ at several $c_{\textrm{MT}}$. 
The data were taken in a magnetic field of 3 Oe perpendicular to the conduction plane.
The demagnetization factor was corrected by means of an ellipsoidal 
approximation.
The slow-cooled condition (filled symbols) gives almost a full Meissner volume
fraction, whereas in the quenched case (open symbols) a slight 
suppression of superconductivity (still above $\sim$ 90\%vol) is found 
in $c_{\textrm{MT}} \geq$ 1\%.
We do not find any trace of magnetic impurity for all the samples,
implying that the present substitution induces non-magnetic disorder.
$T_{\textrm{c}}$ is defined as the intercept of
the extrapolation of the normal and superconducting states.
The superconducting transition in MT5\% becomes broader than that in
MT0\% and MT1\%.
Nevertheless, as shown in the inset of Fig.\ \ref{Fig:kaiT}, 
$T_{\textrm{c}}$ linearly decreases with increasing $c_{\textrm{MT}}$ in both
cooling rates.
The slow-cooled data in the MT substituted $\kappa$-(dET)$_2$Cu(NCS)$_2$ 
also have the same slope as in $\kappa$-(ET)$_2$Cu(NCS)$_2$.

Next we move on to the magnetization curve measurements.
The upper inset of Fig.\ \ref{Fig:Hirr} shows the $M(H)$ curves obtained at 2.0 K
in the slow-cooled condition, where $M$ is the volume magnetization.
The irreversibility field $H_{\textrm{irr}}$, denoted by 
arrows, decreases with increasing $c_{\textrm{MT}}$.
The hysteresis width of the magnetization ($\Delta M$) below $H_{\textrm{irr}}$
at a reduced temperature $T/T_{\textrm{c}}=0.25$ is displayed in the lower inset of
Fig.\ \ref{Fig:Hirr}.
The suppression of $\Delta M$ obviously occurs in high $c_{\textrm{MT}}$.
This behavior reflects its weakened effective vortex pinning.
The reduced temperature ($T/T_{\textrm{c}}$) dependence of $H_{\textrm{irr}}$ is shown in the main 
panel of Fig.\ \ref{Fig:Hirr}. 
As described above, $H_{\textrm{irr}}$ is suppressed with increasing $c_{\textrm{MT}}$.
Thus, the vortex liquid phase (reversible region in $H > H_{\textrm{irr}}$) 
expands.
This also indicates that the effective pinning reduces with $c_{\textrm{MT}}$.
The reduction of the effective pinning will be attributed to decreasing the condensation energy, as
discussed further below.

We turn to deriving the in-plane penetration depth from the
magnetization curves.
Figure 3 shows the semi-logarithmic plot of the reversible magnetization in the 
slow-cooled condition.
An estimation of $\lambda_{\parallel}$ is given from linear slopes
as described below;
for type-II superconductor in $H_{\textrm{c1}} < H_{\textrm{irr}} < H 
\ll H_{\textrm{c2}}$,
the London model\cite{deGennes,891} represents the volume magnetization 
as
$M=-(\alpha\phi_0/32\pi^2 \lambda_{\parallel}^2)\ln(H_{\textrm{c2}}\beta/H),$
\label{Eq:MH}
where $\phi_0$ is the magnetic flux quantum, $\alpha ( = 0.70)$ a correction factor
for vortex-core contribution\cite{891}, and $\beta$ a constant of order unity.
According to this relationship, steeper slope in the $M$ vs. $\ln H$ plot 
gives shorter $\lambda_{\parallel}$.
The solid lines shown in Fig. \ref{Fig:MlnH} exhibit linear regions fitted to this 
relation between $H_{\textrm{irr}}$ and $H_{c2}$. 
As can be seen, the slope of the lines decreases with increasing temperature, so that 
$\lambda_{\parallel}$ ordinarily increases with $T$.

The temperature dependence of $\lambda_{\parallel}$ evaluated above
is shown in Fig. \ref{Fig:lambda}.
Although there is no cooling rate dependence of $\lambda_{\parallel}(T)$ within 
experimental accuracy, 
a marked enhancement of $\lambda_{\parallel}$ with increasing 
$c_{\textrm{MT}}$ is recognized.

In order to estimate $\lambda_{\parallel}(0)$, we apply an
extrapolation to 0 K by using a $d$-wave model\cite{WonMaki,20}:
$\lambda_{\parallel}(T)=\lambda_{\parallel}(0)[1-at-bt^3]^{-0.5}$, where 
$\lambda_{\parallel}(0)$, $a$, and $b$ are fitting parameters.
The solid curves in Fig. \ref{Fig:lambda} are obtained by fitting to the 
slow-cooled data (filled symbols) below 
$T/T_{\textrm{c}} < 0.7$ with $\lambda_{\parallel}(0)=$ 430, 445, and 465
nm, $a=$ 0.12, 0.69, and 0.76, and $b=$ 1.32, 0.41, and  0.56 for MT0\%, 
MT1\%, and MT5\%, respectively.
The parameters $a$ and $b$ are in the same order of magnitude as the
theoretically expected ones ($a=0.6478$ and $b=0.276$)\cite{20}, except for
$b$ in MT0\%.
On the other hand, as overlaid in Fig. \ref{Fig:lambda}, an $s$-wave 
(local clean) model with $\lambda_{\parallel}(0)=435\ $ nm (broken curve) 
also seems to be agreeable to the experimental data in MT0\%.
In the following, we adopt $\lambda_{\parallel}(0)$ obtained from 
the $d$-wave model fits, although
we cannot assert that $\lambda_{\parallel}(T)$ truly 
reflects the unconventional symmetry within the present study.

Before discussing the variation of $\lambda_{\parallel}(0)$ by the MT substitution, 
we mention an intrinsic impurity effect on the penetration depth of the pristine samples.
As in our previous report,\cite{1151} 
$\lambda_{\parallel}(0)$ in $\kappa$-(ET)$_2X$
quantitatively reflects its cleanness;
on the basis of the local clean BCS approximation, 
$\lambda_{\parallel}(0)$ is described as
\begin{equation}
\lambda_{\parallel}(0)=\lambda_{\textrm{L}}(0)(1+\xi_0/l_{\parallel})^{0.5},\label{Eq:London}
\end{equation}
where $\lambda_{\textrm{L}}(0) [=cm^*/4\pi en_s(0)]$ and $\xi_0 (=a \hbar
v_{\textrm{F}}/k_{\textrm{B}}T_{\textrm{c}}$) are the London penetration
depth and coherence length for an ideally pure sample, respectively, 
and $c$ is the light velocity, $m^*$ the effective mass,
$n_s(0)$ the carrier density, $v_{\textrm{F}}$ the Fermi velocity, and $a$ = 0.18.
This indicates that $\lambda_{\parallel}(0)$ becomes long
with decreasing $l_{\parallel}$.
We have observed $\lambda_{\parallel}(0)=430$ nm in 
$\kappa$-(ET)$_2$Cu(NCS)$_2$, which is almost the same with the 
calculated value of $\lambda_{\textrm{L}}(0)=410$ nm.
Taking account of eq. (\ref{Eq:London}), this is due to its long 
mean free path, $l_{\parallel} \sim 150$ nm.\cite{141} 
On the other hand, $\lambda_{\parallel}(0)=570$ nm obtained in $\kappa$-(ET)$_2$Cu[N(CN)$_2$]Br,
which is relatively larger than $\lambda_{\textrm{L}}(0)= 400$ nm,
is attributed to short $l_{\parallel}$ ($\sim$ 40 nm), 
implying an intrinsically dirtier system.
Moreover, its fast-cooling process induces much more disorder, resulting in a 
further enhancement of $\lambda_{\parallel}(0)$ with decreasing 
$l_{\parallel}$ as displayed in Fig. \ref{Fig:London} (filled diamonds)\cite{1151}.
This can also be interpreted on the basis of eq. 
(\ref{Eq:London}) (small circles in Fig. \ref{Fig:London}).
Thus, the local clean model adequately describes the absolute value
of $\lambda_{\parallel}(0)$ in these organics.

The MT molecules, most probably working as impurity,
will give rise to scattering quasiparticles and decreasing $l_{\parallel}$.
Consequently $\lambda_{\parallel}(0)$ is varied via $l_{\parallel}$,
whereas $\lambda_{\textrm{L}}(0)$ would not be influenced by the MT substitution.
Very recently, the scattering time $\tau_{\parallel}$ is 
systematically studied in the present system by means of the de 
Haas-van Alphen oscillation effect.\cite{Oizumi} 
Since $l_{\parallel}=v_{\textrm{F}} \tau_{\parallel}$, one can obtain 
$l_{\parallel}$, only if $v_{\textrm{F}}$ is independent of 
$c_{\textrm{MT}}$.
In this experiment, $\tau_{\parallel}=3$ ps in MT0\%, giving 
$l_{\parallel}=135$ nm by applying $v_{\textrm{F}}=4.5\times10^6$ cm/s, 
varies into $\tau_{\parallel}=1.6$ ps (so $l_{\parallel}=$ 72 nm) 
in MT1\% and 1 ps (45 nm) in MT5\%,
whereas $m^*$ does not alter at all.\cite{Oizumi}
As a function of $l_{\parallel}$, $\lambda_{\parallel}(0)$ is plotted in
Fig. \ref{Fig:London} (filled symbols).
We note that the local clean limit condition holds for the present case
as displayed in Table \ref{t1}, where
the in-plane coherence length at 0 K, $\xi_{\parallel}(0)$,
is estimated from the form 
$\xi_{\parallel}^{-1}=\xi_0^{-1}+l_{\parallel}^{-1}$.

Now, by using $l_{\parallel}$ obtained from $\tau_{\parallel}$, one can 
evaluate $\lambda_{\parallel}(0)$ according to eq. (\ref{Eq:London})  
with a fitting parameter $\lambda_{\textrm{L}}(0)=420$ nm, which is 
shown in Fig. \ref{Fig:London} (small open circles).
As described in this figure, the calculated $\lambda_{\parallel}(0)$ 
is fairly consistent with the experimental data.
Thus the absolute value of $\lambda_{\parallel}(0)$ is quantitatively 
explained from the view of the impurity effect in the same manner 
in fast-cooling of $\kappa$-(ET)$_2$Cu[N(CN)$_2$]Br.

\begin{table}[tb]
\caption{Superconducting parameters in the MT substituted 
$\kappa$-(ET)$_2$Cu(NCS)$_2$ in the slow-cooled condition.}
\label{t1}
\begin{tabular}{ccccccc}
\hline
 $c_{\textrm{MT}}$ & $T_{\textrm{c}}$ & $\lambda_{\parallel}(0)$ &  
 $\l_{\parallel}$ & $\xi_{\parallel}(0) $ & $U_p$ \\ 
 (\%) & (K) & (nm) & (nm) & (nm) & ($10^{-13}$ J/m$^3$) \\
\hline
 0 & 9.2 & 430 & 135 & 6.4 &  3.7 \\
 1 & 9.0 & 445 & 72 & 6.5 &  3.5 \\
 5 & 7.9 & 465 & 45 & 6.6 & 3.1 \\
\hline
\end{tabular}
\end{table}

We finally discuss the effective vortex pinning in terms of
the superconducting condensation energy.
The distribution of impurity sites originating from the MT substitution 
is very likely to be uniform in the whole bulk.
The size of them is presumably in a molecular scale (a 
few \AA) which is much smaller than the vortex core diameter,
$2\xi_{\parallel} \sim 13$ nm.
Thus, the introduction of MT molecules does not contribute to vortex 
pinning, but curiously reduces the effective pinning as 
seen in the suppression of $\Delta M$ with $c_{\textrm{MT}}$.
Such behavior is very similar to the case of fast-cooling in 
$\kappa$-(ET)$_2$Cu[N(CN)$_2$]Br,
in which $\Delta M$ critically reduces with cooling rate.
Although a disorder-domain induced pinning model has been 
proposed,\cite{1030} this is not applicable to the present case.
As a reasonable explanation on the effective pinning
reduced by both methods, the MT substitution and fast-cooling,
which may originate from a similar mechanism, 
we focus on the superconducting condensation energy.
Namely, the pinning potential for a normal-superconducting 
boundary ($U_p$) is described as the condensation energy of the vortex 
core, which is given by $U_p=(1/2)\mu_0H_{\textrm{c}}^2 \cdot \pi 
\xi_{\parallel}^2=\phi_0^2/(16\mu_0\pi\lambda_{\parallel}^2)$, 
where $\mu_0H_{\textrm{c}}$ is the thermodynamic critical field expressed as 
$\mu_0H_{\textrm{c}}=\phi_0/(2\sqrt2 \pi \lambda_{\parallel} \xi_{\parallel})$.
We note that $U_p$ depends on only $\lambda_{\parallel}$,
and in addition, smaller $U_p$ will cause weaker effective pinning.
As shown in Table \ref{t1}, $U_p$ decreases with 
$c_{\textrm{MT}}$, which is consistent with the present reduction of the
effective pinning. 
The similar calculation in $\kappa$-(ET)$_2$Cu[N(CN)$_2$]Br leads to
$U_p=2.1$ and $1.4 \times 10 ^{-13}$ J/m$^3$ in the slow-cooled and 
quenched conditions, respectively. 
This is the same trend with the MT substitution.
For further quantitative discussion on the vortex pinning, however, the
origin of vortex pinning should be clarified in the future.

In conclusion, we have observed an impurity effect 
by molecular substitution.
The substituted MT molecules clearly work as impurity giving rise to 
scattering quasiparticles, and so $T_{\textrm{c}}$ monotonically decreases.
The enhancement of $\lambda_{\parallel}(0)$, which is adequately
interpreted as the impurity effect on the basis of the local clean limit
approximation, results in the decrease in the condensation energy.
It can also account for the reduction of the effective pinning.

\section*{Acknowledgment}
This research was partly supported by the Ministry of Education,
Science, Sports and Culture, Grant-in-Aid for Encouragement of Young
Scientists (Grant No. 17750119), and for Scientific Research (B) (Grant 
No. 17340099).

\end{document}